\documentclass[showpacs,amssymb,floatfix]{revtex4}[14pt]
\usepackage{graphicx}
\usepackage{dcolumn}
\usepackage{graphics}
\usepackage{epstopdf}
\usepackage{pdfpages}
\usepackage{bm}
\begin{document}
\title{Interference of Laguerre--Gaussian beams for reflection by dielectric slab}
\author{Konstantin N. Pichugin and Almas F. Sadreev}
\affiliation{$^1$ Kirensky Institute of Physics, Federal Research Center KSC SB RAS, 660036
Krasnoyarsk, Russia}
\date{\today}

\begin{abstract}
 We study reflection of TE Laguerre-Gaussian  light beam
 by dielectric slab and show that the Goos-H\"{a}nchen  and  the  Imbert-Federov  shifts
show resonant behavior following to the behavior of the reflection. Moreover the Imbert-Federov linear
and the Goos-H\"{a}nchen angular shifts strongly   depend on the orbital angular momentum $m$.
Due to destructive interference of two beams reflected from upper and down interfaces
of the slab profile of the reflected light beam acquires structure
which distinctively displays an amount of $m$.
 \end{abstract}
%\pacs{42.25.Fx,2.60.Da,42.79.Dj}
 \maketitle

 %% activate for two-column option
 \section{Introduction}

%from Aiello&Woerdman OAM shifts
It is well established that a bounded beam upon reflection and transmission on a planar interface
differs in propagation with plane waves due to diffraction corrections. This may manifest as beam
shifts with respect to the geometric optics prediction when reflected or refracted.
The more dominant shifts are the Goos-H\"{a}nchen (GH) shift in which the beam is
displaced  parallel  to  the  plane  of  incidence \cite{Goos},  and  the  Imbert-Federov (IF) shift
in which the shift is  perpendicular \cite{Imbert}. Moreover, it has been shown
that each of these two beam shifts can be separated into a spatial and an angular shift.
The main distinction between spatial and angular shifts is the enhancement of the latter
with the propagation of the beam \cite{Aiello}.

\begin{figure}[ht]
\includegraphics[width=8cm,clip=]{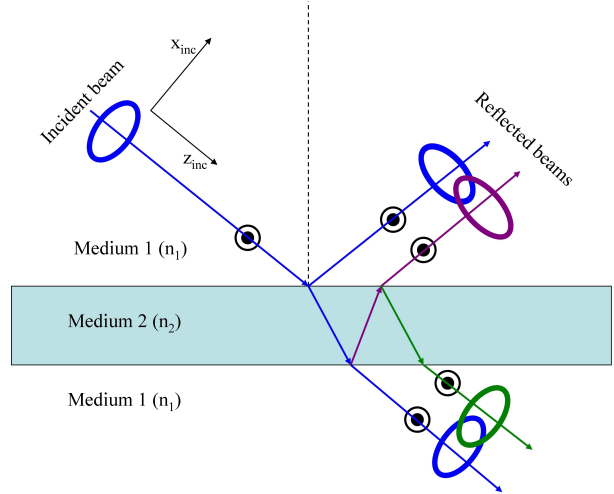}
\caption{(Color online) Gaussian beam reflected and
refracted on a dielectric slab with thickness $d$.} \label{fig1}
\end{figure}

In the present letter we consider transmission and reflection of
Laguerre-Gauss (LG) beams carrying orbital angular momentum (OAM)
by dielectric slab. Specifically the effect of orbital angular
momentum (OAM) on GH and IF shifts was considered for reflection
of the LG beam from an interface
\cite{Allen,Fedoseyev2008,Bliokh2009,Merano2009,MeranoPRA,Aiello2012,Merano2013}.
Along with that reflection of Gaussian beams by dielectric slabs
was considered to show negative shifts and resonant behavior of GH
and IF shifts \cite{Riesz,Li2003,Wang2005,Wen2017}. First reflection and transmission of LG beam in dielectric slab
was considered by Li {\it et al} \cite{Li}.  The intensity distributions of the reflected and transmitted beams were presented which
unambiguously show interference of two LG beams reflected from different interfaces and effects of vortices.
In the present paper we develop these studies  for reflection of LG beam by dielectric slab as
sketched in Fig. \ref{fig1}. Owing to double reflection by two
planes of the slab gives rise to an interference of two LG beams
that in turn results in strong dependence of the GH and IF shifts
on an amount of OAM. The most bright effect of such interference
is a resonant behavior of GH and IF shifts as dependent on the angle of
incidence and frequency of the incident beam.
%--------------------------------------------------------------------------------------------Fig.1
%From P.K. Swain,N. Goswami, and A. Saha, "RSP
%resonance enhanced spatial and angular Goos–Hanchen shift and
%Imbert–Fedorov shift for Gaussian beam, Laguerre–Gaussian beam and
%Bessel beam", Opt. Comm. \textbf{382} (2017) 1--6.

\section{The method}
The complex electric field amplitude of the reflected
and transmitted beams is calculated using the angular spectrum
method \cite{McGuirk} and paraxial approximation. Following to
Okuda and Sasada \cite{Okuda} we do not consider
cross-polarization effects \cite{Kohazi,Bliokh2} because we choose
the incident beam to have a small paraxial parameter
$1/kw_0=0.01$. It is worthy to notify paper by Ou {\it el}  where reflection of LG beam by dielectric surface was considered beyond the
paraxial approximation by a full Teylor expansion \cite{Ou}.
Then, the TE or TM component of incident electric field has form
%-------------------------------------------------------------------------------------------------Eq. (1)
\begin{equation}\label{inc}
E_{mp}(x,y,z)=u_{mp}(\rho_{inc}, \phi_{inc},z_{inc})e^{ikz_{inc}}
\end{equation}
where $u_{mp}(\rho,\phi,z)$ is the LG-mode envelope function with the azimuthal and radial
indices $m$ and $p$ respectively:
%-------------------------------------------------------------------------------------------------Eq. (2)
\begin{eqnarray}\label{LG}
&
u_{mp}(\rho,\phi,z)=C_{mp}(-1)^p\frac{w_0}{w(z)}\left(\frac{\sqrt{2}\rho}{w(z)}\right)^{|m|}
L_m^{|p|}\left(\frac{2\rho^2}{w(z)^2}\right)e^{im\phi}
&\nonumber\\
&\exp\left[-\frac{\rho^2}{w(z)^2}+\frac{ik\rho^2}{2R(z)}-i(2p+|m|+1)\psi(z)\right]&
\end{eqnarray}
with $C_{mp}=\sqrt{\frac{2p!}{\pi(p+|m|)!}}$, where $k=k_0n_1$,
%-------------------------------------------------------------------------------------------Eqs. (3)-(5)
\begin{eqnarray}\label{ww}
&w(z)=w_0\sqrt{1+\left(\frac{2z}{kw_0^2}\right)^2},&\\
&R(z)=z\left[1+\left(\frac{kw_0^2}{2z}\right)^2\right],&\\
&\psi(z)=arctan\left(\frac{2z}{kw_0^2}\right).&
\end{eqnarray}
%--------------------------------------------------------------------------------------------Fig.1
\begin{figure}[ht]
\includegraphics[width=4.5cm,clip=]{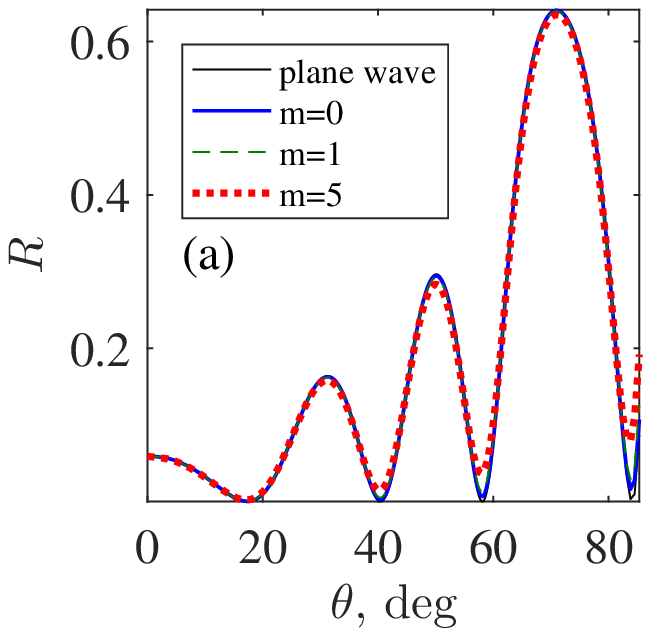}
\includegraphics[width=4.5cm,clip=]{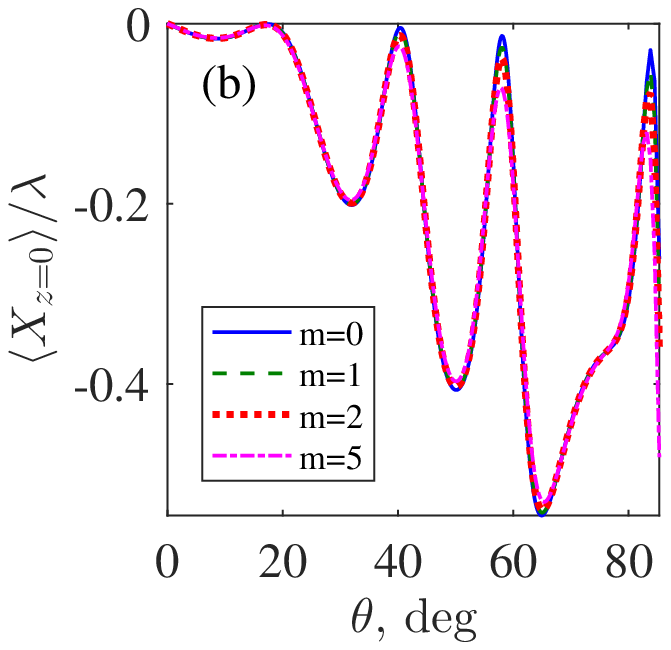}
\includegraphics[width=4.5cm,clip=]{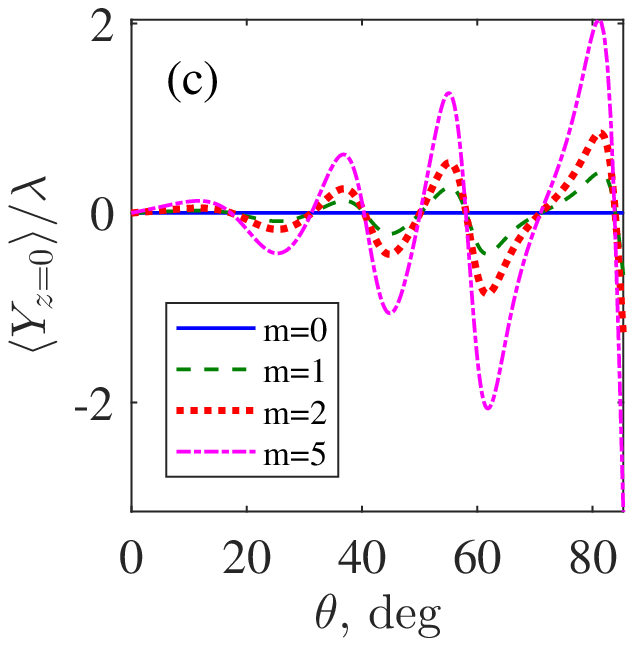}
\includegraphics[width=4.5cm,clip=]{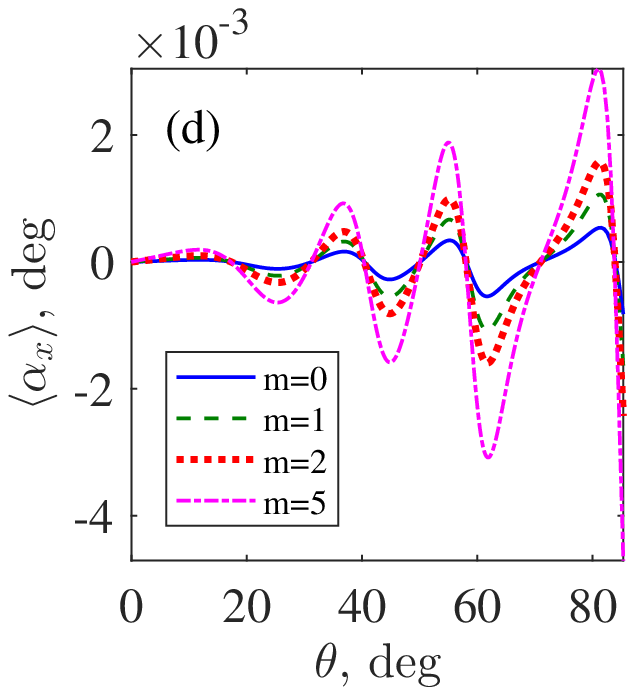}
\caption{(Color online) (a) The reflection of the LG beam with
different OAM from dielectric slab of thickness
$d=7.96\lambda$ and permittivity $\varepsilon=2$
(glass) vs the angle of incidence of the beam. (b)
Goss-H\"{a}nchen, (c) Imbert-Fedorov shifts, and (d) angular
deviation for reflection.} \label{fig2}
\end{figure}

The angular spectral amplitude of the incident LG beam over the
dielectric interface is given by Fourier transformation as
%-------------------------------------------------------------------------------------------(6)
\begin{equation}\label{Four}
E_{mp}(k_x,k_y)=\int dxdyE_{mp}(x,y,z=0)e^{-ik_xx-ik_yy}.
\end{equation}
Multiplying Eq. (\ref{Four}) by the amplitude reflectance
$R(k_x,k_y)$ and the propagation phase factor $\exp(ik_zz)$
\cite{Okuda}, the amplitude of the reflected LG beam is obtained
through inverse Fourier transformation
%--------------------------------------------------------------------------------------------(7)
\begin{eqnarray}\label{refl}
&E_{mp}^{(r)}(x,y,z)=\frac{1}{(2\pi)^2}\int dk_xdk_y R(k_x,k_y)E_{mp}(k_x,k_y)&\nonumber\\
&\exp[i\sqrt{k^2-k_x^2-k_y^2}z+i(k_xx+k_yy)].&
\end{eqnarray}
The central angular component of the incident paraxial beam is
$(k_x,k_y,k_z)=(k\sin\theta, 0, k\cos\theta)$, the other components
are mainly confined within
$$|k_x/k-\sin\theta|,  |k_y/k|,  |kz/k-\cos\theta|\leq (kw_0)^{-1}$$
around the central component, where $(kw_0)^{-1}=0.01$ is the
paraxial parameter \cite{Lax}.
For numerical integration we use
$|k_x-\sin\theta|<7/w_0\cos\theta$ and $|k_y-\sin\theta|<7/w_0$.
Using transfer matrix method to calculate reflection and
transmission coefficients through a slab we get
%--------------------------------------------------------------------------------------------(8)
\begin{eqnarray}\label{RT}
&R=r\left(1-\frac{k_{z2}}{k_{z1}}\frac{\exp(2idk_{z2})t^2}{1-r^2\exp(2idk_{z2})}\right),&\\
&T=\frac{k_{z2}}{k_{z1}}\frac{\exp(idk_{z2})t^2}{1-r^2\exp(2idk_{z2})}.&
\end{eqnarray}
Here $r,t$ are Fresnel coefficients for TE or TM modes
%--------------------------------------------------------------------------------------------(9)
\begin{eqnarray}\label{r}
&r_s(k_x,k_y)=\frac{k_{z1}-k_{z2}}{k_{z1}+k_{z2}},&\\
&t_s(k_x,k_y)=2\frac{k_{z1}}{k_{z1}+k_{z2}},&\\
&r_p(k_x,k_y)=\frac{n_2^2k_{z1}-n_1^2k_{z2}}{n_2^2k_{z1}+n_1^2k_{z2}},&\\
&t_p(k_x,k_y)=2\frac{n_2^2k_{z1}}{n_2^2k_{z1}+n_1^2k_{z2}}.&
\end{eqnarray}
with $k_{zi}=\sqrt{k_i^2-k_x^2-k_y^2}$ and $k_i=k_0n_i$.
\section{The results}
Fig. \ref{fig2} (a) shows reflection of the LG beam with different
OAM from dielectric slab versus the angle of incidence. One can
see that the amount of OAM not strongly effect the reflection and
the Goss-H\"{a}nchen shift. The obvious effect is related to
disappearance of zeros of the reflection. However as Fig.
\ref{fig2} (c) and (d) show the  Imbert-Fedorov shift and
longitudinal angular shift are considerably depend on the OAM  for
reflection from the dielectric slab. The transversal angular shift
is negligibly small of order $10^{-8}$ deg and is not shown.

\begin{figure}[ht]
\includegraphics[width=4.5cm,clip=]{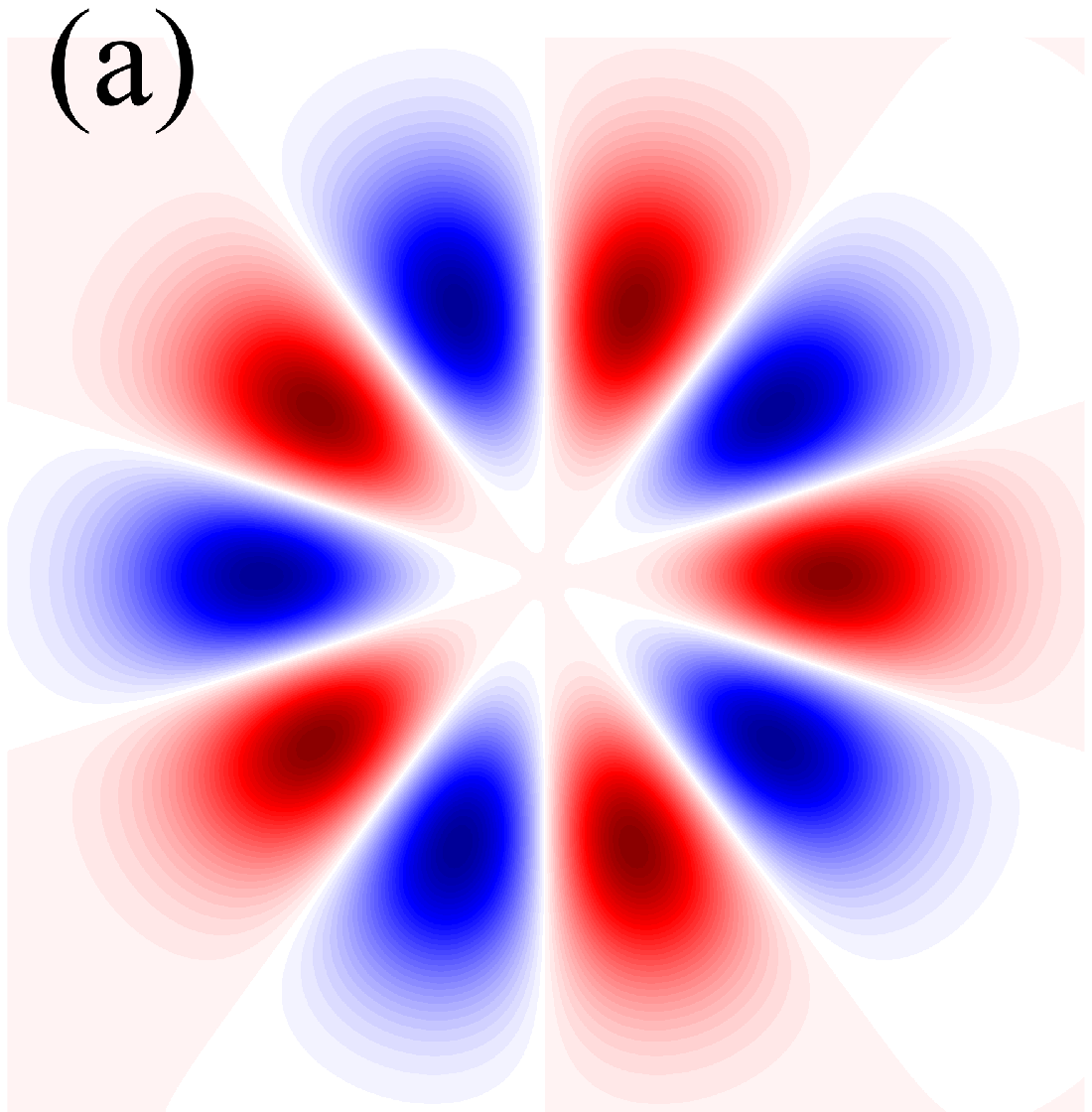}
\includegraphics[width=4.5cm,clip=]{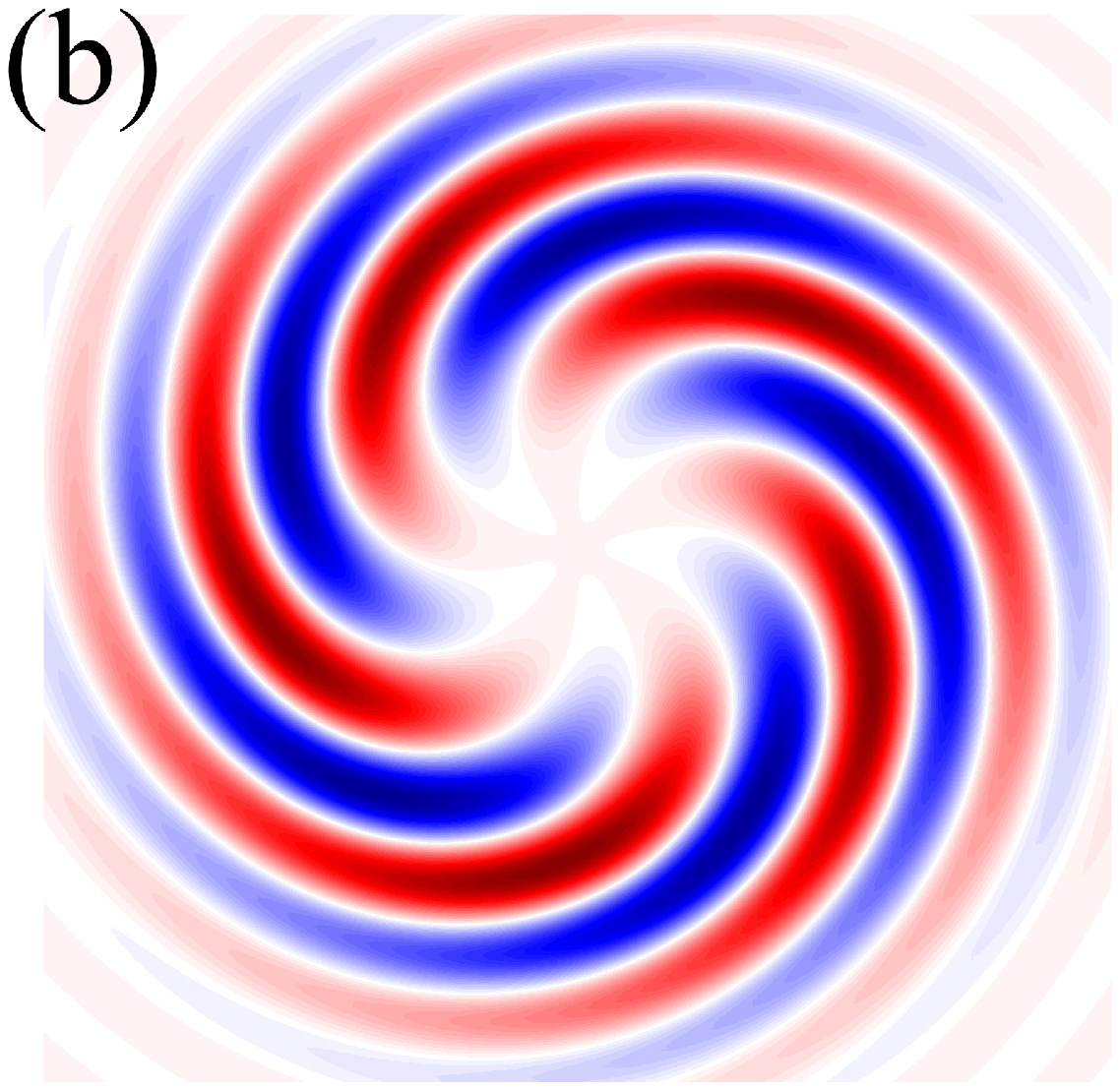}
\caption{(Color online) Real parts of the incident LG beam (\ref{LG}) at $z=0$ (a) and $z=100$ (b).}
\label{fig3}
\end{figure}

As seen from Fig. \ref{fig1} for reflection of LG beam from the dielectric slab there is interference
of, at least,
two beams with destructive interference of beams with opposite OAMs. one can expect that total beam
can have complicated
structure in plane normal to the line of reflected beam. Fig. \ref{fig3} show
the structure of the incident LG beam, its real part  at $z=0$
where a waist of the beam is minimal and at
$z=100$ where the real part of function (\ref{LG}) demonstrates its spiralled structure
because of difference in velocities of the beam at the center and the periphery.
%--------------------------------------------------------------------------------------------Fig.3

\begin{figure}[ht]
\includegraphics[width=8cm,clip=]{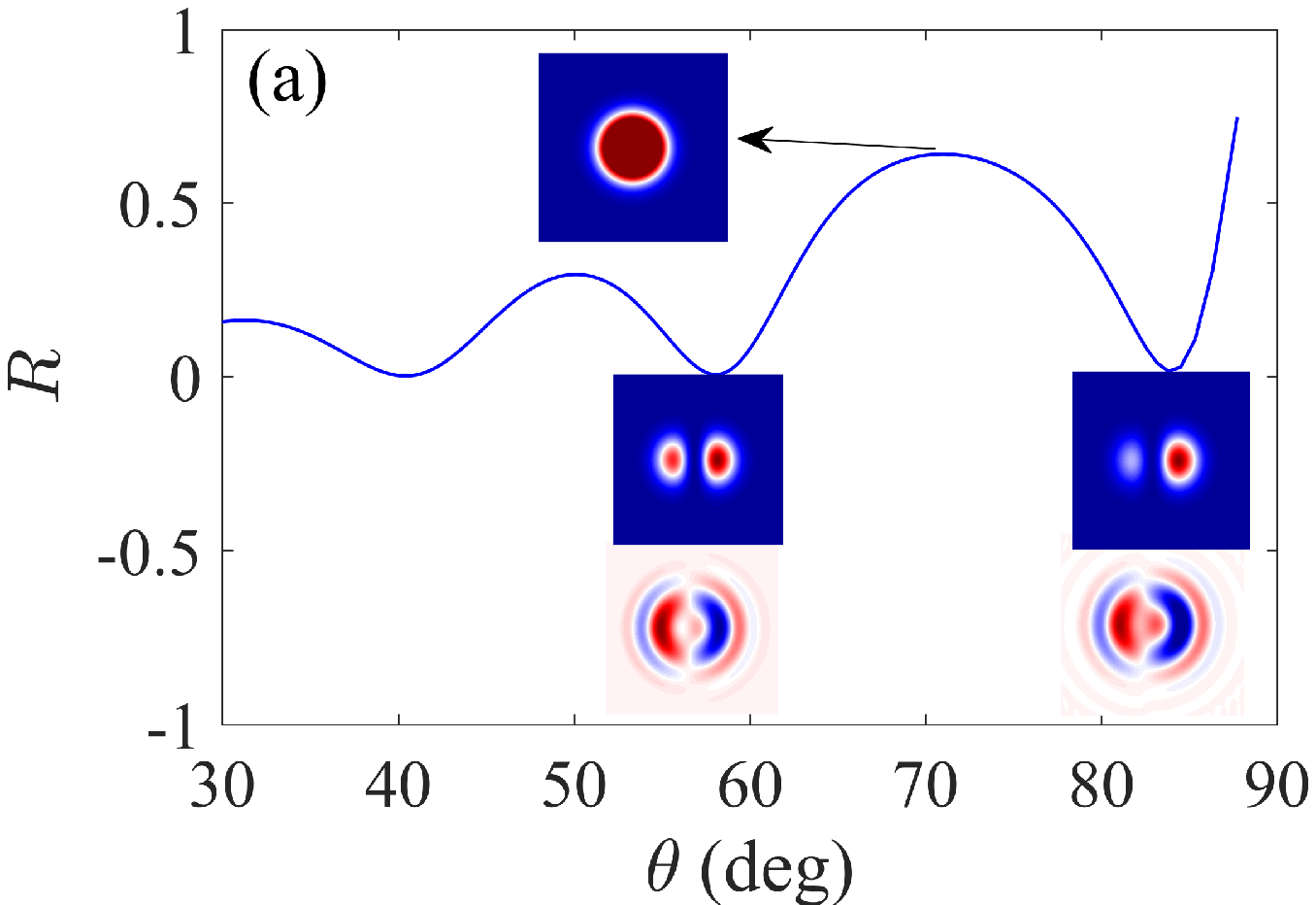}
\includegraphics[width=8cm,clip=]{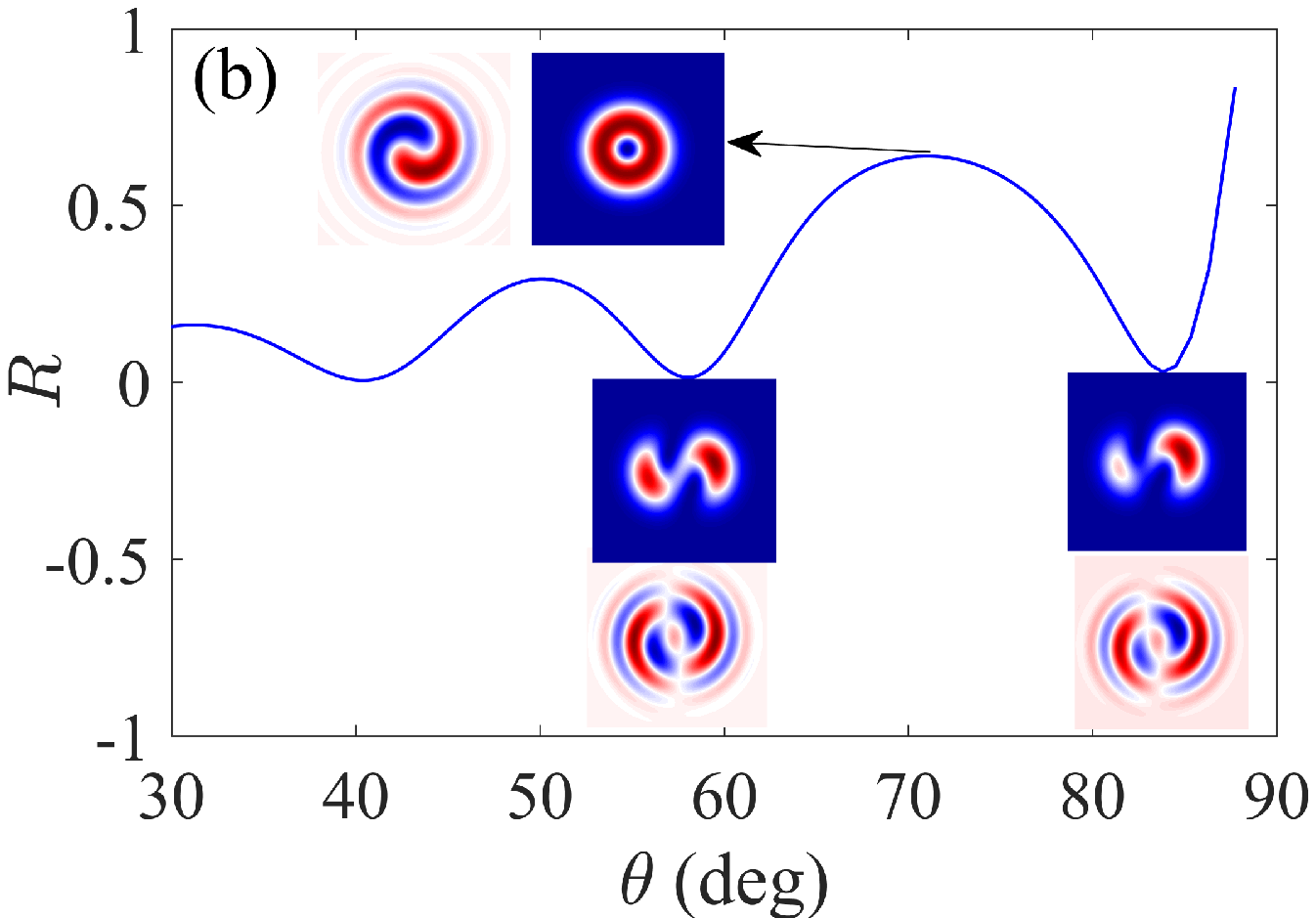}
\includegraphics[width=8cm,clip=]{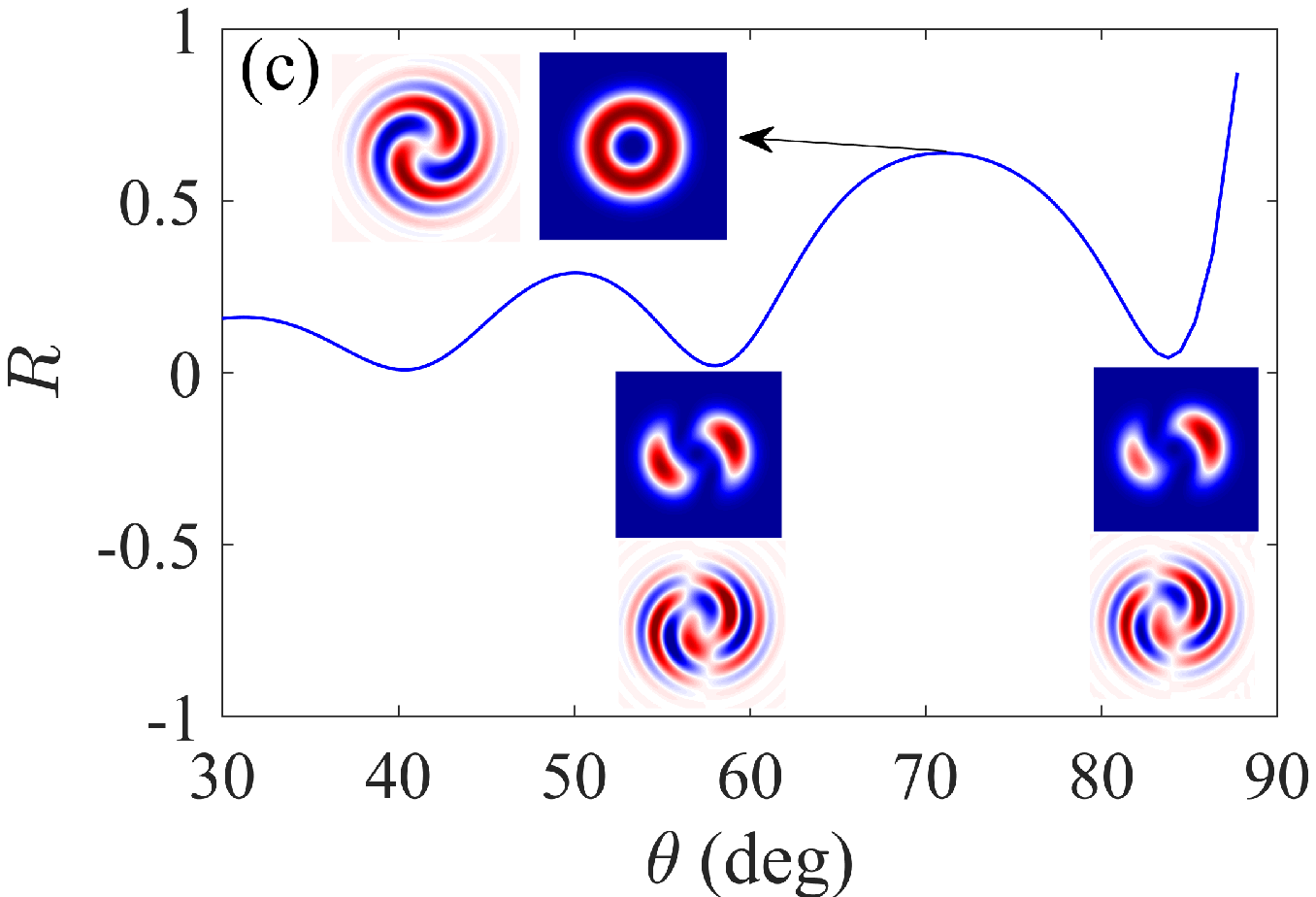}
\includegraphics[width=8cm,clip=]{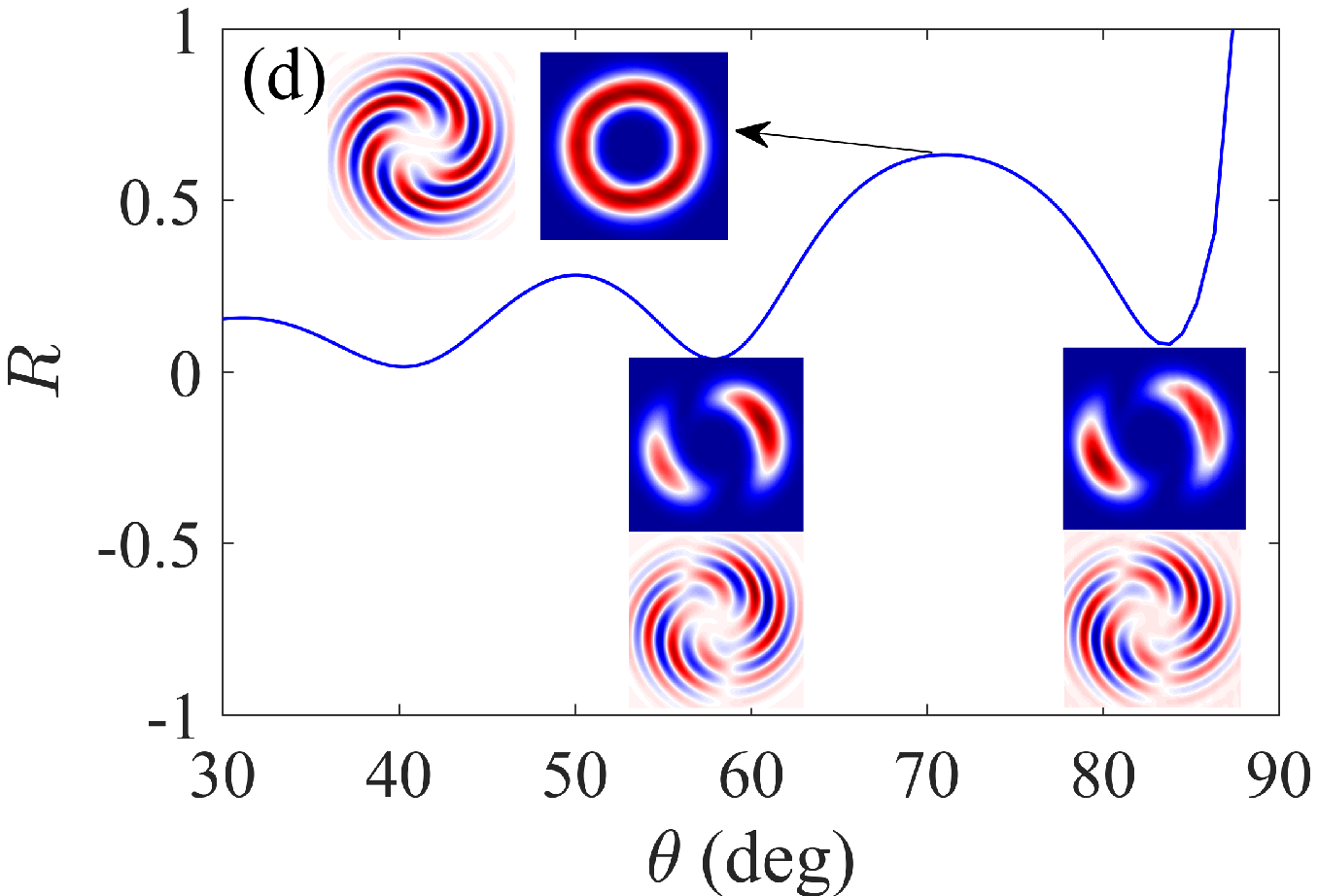}
\caption{(Color online) Real parts and modulus of the function (\ref{refl}) for reflection of LG
beams  by slab with the parameters
given in Fig. \ref{fig2} for (a) $m=0$, (b) $m=1$, (c) $m=2$, and (d) $m=5$.}
\label{fig4}
\end{figure}

In series of Figures in Fig. \ref{fig4} we show evolution of the intensity
and amplitude of the interfered LG beams with different OAM at those angles of incidence which
correspond to minima and maxima of reflection.
%--------------------------------------------------------------------------------------------Fig.4
Comparison of Fig. \ref{fig4} (b) to Fig. \ref{fig3} (b) shows that the LG beam after reflection
changes a chirality because of inversion of velocities. Moreover the structure of real part of the
beam also changes significantly and distinctively reflects the amount of OAM in the LG beam.

\begin{figure}[ht]
\includegraphics[width=4.5cm,clip=]{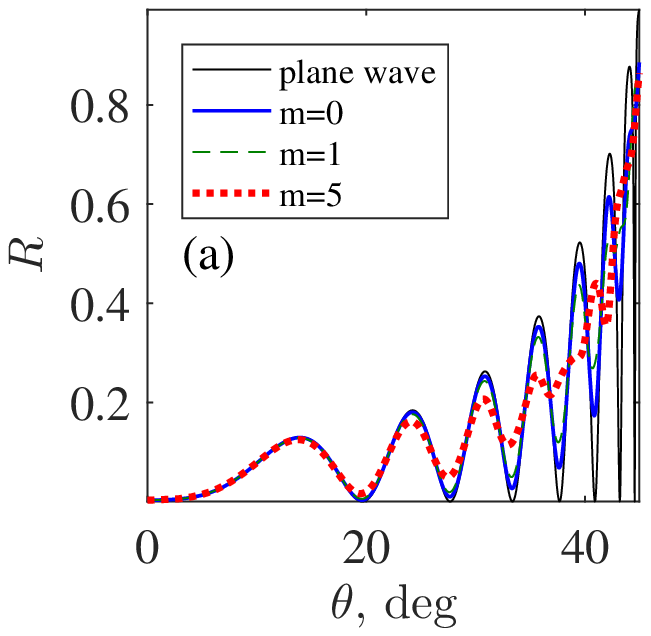}
\includegraphics[width=4.5cm,clip=]{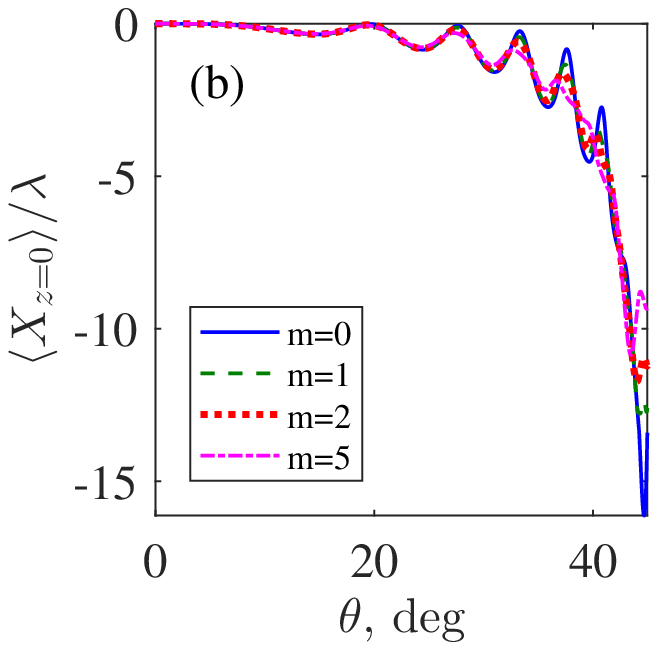}
\includegraphics[width=4.5cm,clip=]{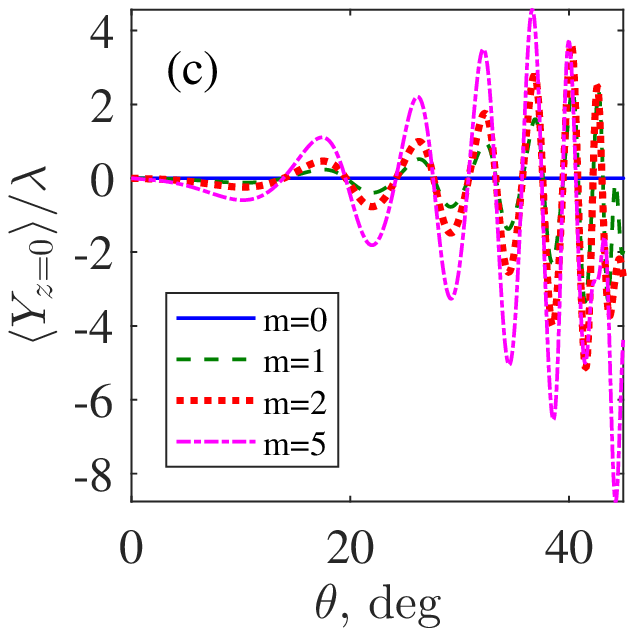}
\includegraphics[width=4.5cm,clip=]{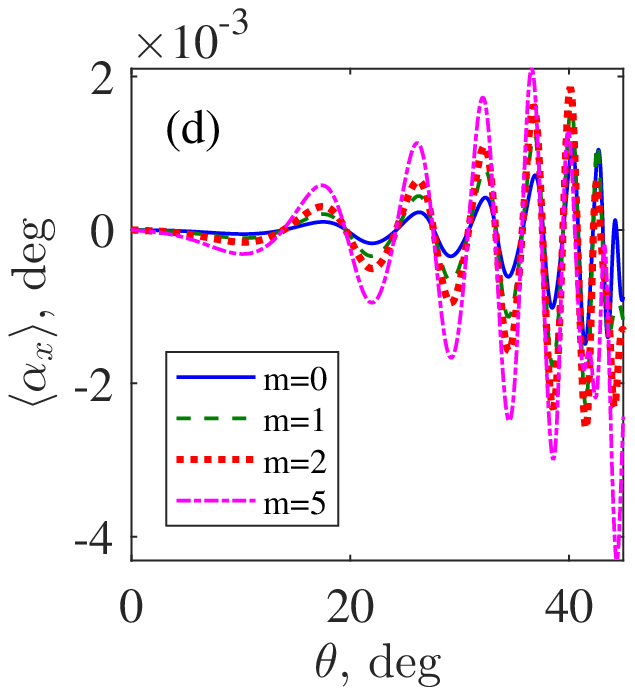}
\caption{((Color online) (a) The reflection of the LG beam with
different OAM from air slot of thickness $d=5.63\lambda$
in air vs the angle of incidence of the beam. (b)
Goss-H\"{a}nchen, (c) Imbert-Fedorov shifts, and (d) angular
deviation for reflection.} \label{fig5}
\end{figure}

Till now we considered the reflection of the LG beams from slab in
air. In Fig. \ref{fig5} we present results for GH and
IF shifts for the opposite case of air slot in glass.
%--------------------------------------------------------------------------------------------Fig.4

\section{Discussion and summary}
Bliokh {\it et al} \cite{Bliokh2009} predicted vortex-induced GH
shift related to the angular IG angular shift for reflection of
the LG beam from an interface separating two dielectric media.
Similar phenomena take place for reflection from dielectric slab
with two interfaces however the GH and IF shifts take resonant
behavior with the angle of incidence as shown in Fig. \ref{fig3}.
As one can expect an amount of the IF shifts strongly depend on
the value of OAM $m$, i.e., on a vorticity of LG beams. It is
clear that sign of the IF shifts depends on the sign of $m$.
Second, for reflection from the dielectric slab the reflection and
GH shift both weakly depend on OAM, while for reflection from air
slot in dielectric medium the amount of OAM of LG beam affect
resonant behavior of the reflection and GH shift as shown in Fig.
\ref{fig5} (a) and (b). Third, reflection of LB beam from slab and
slot both gives rise to interference of neighbor LG beams that in
turn results in structures of reflected light shown in Fig.
\ref{fig4}.

We acknowledge discussions with Dmittrii Maksimov and Jayachandra Bingi. This  work  was
supported  RFBR grant 17-52-45072.

\end{document}